\documentclass[12pt]{article}

\usepackage{graphicx,amssymb,amsmath,empheq,color}
\usepackage{simpler-wick}
\usepackage[export]{adjustbox}
\usepackage{hyperref}
\hypersetup{colorlinks = true, linkcolor=red, citecolor=blue, urlcolor=blue}
\usepackage{url}
\usepackage{caption}
\usepackage{subcaption}

\makeatletter

\newlength{\fighskip} \fighskip=2pt
\newlength{\figvskip} \figvskip=3pt

\newcommand*{\figbox}[2]{{
  \def\figscale{#1}
  \def\arraystretch{0.8}
  \arraycolsep=0pt
  \begin{array}{c}
    \vbox{\vskip\figscale\figvskip
      \hbox{\hskip\figscale\fighskip
        \includegraphics[scale=\figscale]{#2}}}
  \end{array}}}

\newcommand*{\wideboxed}[1]{\setlength{\fboxsep}{1ex}%
  \fbox{\m@th$\displaystyle#1$}}

\def\ubrace#1_#2{%
  \underbrace{#1}_{\hb@xt@\z@{\hss$\scriptstyle#2$\hss}}}

\newcommand{\blangle}{\bigl\langle}
\newcommand{\brangle}{\bigr\rangle}
\newcommand{\dlangle}{\langle\kern-1.5pt\langle}
\newcommand{\drangle}{\rangle\kern-1.5pt\rangle}
\newcommand{\bdlangle}{\blangle\kern-3pt\blangle}
\newcommand{\bdrangle}{\brangle\kern-3pt\brangle}

\newcommand*{\bra}[1]{\langle#1|}
\newcommand*{\ket}[1]{|#1\rangle}

\newcommand{\vep}{\varepsilon}
\newcommand{\vp}{\varphi}

\newcommand{\be}{\begin{equation}}
\newcommand{\ee}{\end{equation}}
\newcommand{\p}{\partial}

\newcommand{\calA}{\mathcal{A}}

\newcommand{\calF}{\mathcal{F}}

\newcommand{\calH}{\mathcal{H}}

\newcommand{\calJ}{\mathcal{J}}
\newcommand{\calK}{\mathcal{K}}

\newcommand{\calS}{\mathcal{S}}

\newcommand{\calZ}{\mathcal{Z}}

\newcommand{\TFD}{\mathrm{TFD}}

\newcommand*{\trans}{{\mathpalette\@trans{}}}
\newcommand*{\@trans}[2]{\raisebox{\depth}{$\m@th#1\intercal$}}

\DeclareMathOperator{\Tr}{Tr}

\DeclareMathOperator{\TT}{\mathbf{T}}

\DeclareMathOperator{\Sch}{Sch}
\DeclareMathOperator{\CRT}{CRT}

\definecolor{DarkGreen}{rgb}{0.0, 0.5, 0.0}

\makeatother

\title{Disentangling the thermofield-double state}

\author{Pouria Dadras\footnote{pdadras@caltech.edu}\\
\normalsize\it California Institute of Technology, Pasadena, CA 91125, U.S.A.\vspace{0.5cm}}

\date{}

\begin{document}

\setcounter{tocdepth}{2}

\maketitle
\begin{abstract}
In this paper, we consider the evolution of the thermofield-double state under the double-traced operator that connects its both sides. We will compute the entanglement entropy of the resulting state using the replica trick for the large N field theory.  To leading order, it can be computed from the two-point function of the theory, where, in CFTs, it is fixed by the symmetries. Due to the exponential decay of the interaction, the entanglement entropy saturates about the thermal time after the interaction is on. Next, we restrict ourselves to one dimension and assume that the theory at strong coupling is effectively described by the Schwarzian action. We then compute the coarse-grained entropy of the resulting state using the four-point function. The equality of the two entropies implies that the double-traced operators in our theory act \emph{coherently}.  In AdS/CFT correspondence  where the thermofield-double state corresponds to a two-sided black hole, the action of a double-traced operator corresponds to shrinking or expanding the black hole in the bulk.
\end{abstract}

\tableofcontents
\newpage
\section{Introduction}\label{sec:introduction}
Soon after Albert Einstein developed the general theory of relativity, an exact solution to Einstein's equation was found by Karl Schwarzschild, known as the Schwarzschild metric \cite{WaldGR,Hawking73}. The metric describes a vacuum solution in  asymptotically flat spacetime with spherical symmetry. However, it exhibits a peculiar behavior at the Schwarzschild radius, where some  metric components become singular. This hypersurface, known as the black hole's event horizon, has the unusual property that, according to an outside observer, it will take an infinite time for an arbitrary object to reach the horizon, while the particle itself only experiences a finite amount of time. In addition, an event horizon is like a one-way membrane; namely, an object which has already passed the horizon is never able to reach out to the outside and ultimately ends up hitting the physical singularity located at the black hole's center.  Nevertheless, the singularity of the black hole's event horizon is not physical and can be removed by choosing an appropriate coordinate system, e.g., the Eddington-Finkelstein (EF) coordinate system. \\
A somewhat surprising fact about the Schwarzschild and EF coordinate systems is that they only describe a portion of the spacetime. In other words, it is possible to opt for a coordinate system, for example, the Kruskal-Szekeres coordinate, that covers the whole spacetime where the geodesics parametrized by the affine parameter either extend to infinity or terminate by hitting a physical singularity. It turns out that the maximally extended spacetime, in addition to the black hole region, also has a white hole region (its time reversal), and they indeed have two sides that are connected by a non-traversable wormhole called the Einstein-Rosen (ER) bridge. Such extended horizons are called the bifurcate horizons. \\
There are also other solutions to Einstein's equation corresponding to the rotating and charged black holes. Such solutions, although more complicated, capture many features of the Schwarzschild black hole. For example, the maximally extended spacetime contains a bifurcate horizon. Indeed, as is pointed out by Racz and Wald  \cite{Racz92,Racz96}, any stationary spacetime which has a ``one-sided black hole'' but no white hole with its Killing vector's orbits to be diffeomorphic to $\mathbb{R}$ can always be locally extended to a spacetime with a bifurcate horizon provided that the horizon's surface gravity is a nonzero constant.   
Such horizons are usually identified by their mass, electric charge, and angular momentum \cite{Israel1, Israel2, carter71}. 
Further developments by Hawking, Carter, and Bardeen \cite{Bardeen73} showed that assuming the spacetime is a solution to  the Einstein's equation and the matter satisfies the dominant energy condition, it is possible to associate a well-defined temperature to black hole's horizon. Moreover, the black hole satisfies the laws of thermodynamics, with its energy being equal to the black hole's mass and entropy proportional to the horizon's area. \\
 A powerful tool to understand the nature of bifurcate horizons is the AdS/CFT correspondence \cite{Malda98, Witten98} which conjectures that roughly speaking, a gravity theory in the bulk of Anti-de Sitter (AdS) spacetime, which may include black holes as well, is equivalent to a specific conformal field theory living on the boundary. As a part of the dictionary between the bulk and the boundary \cite{Malda01}, the existence of a two-sided black hole in the bulk is dual to the case where two copies of the boundary theory act on each side of the thermofield-double state $(\TFD)$ prepared as the initial state \footnote{The correspondence between thermofield-double state and a two-sided black hole was first observed by Israel in asymptotically flat spacetime \cite{Israel3}.} :
 \be
 \ket{\TFD} = \frac{1}{\calZ^{1/2}} ~ \sum_{n} e^{-\frac{\beta E_n}{2}} \ket{E^*_n}_L\ket{E_n}_R,
 \ee
 where $\ket{E_n}_R$s are the energy eigenstates \footnote{For rotating black holes, the states are labeled by energy and angular momentum eigenvalues.} of the right Hamiltonian, $\ket{E^*_n}_L$ are the eigenstates of $H^*_R$, and $\beta$ is the black hole's inverse temperature. Note that TFD is invariant under the symmetry generated by $1 \otimes H_R- H^*_R \otimes 1 \equiv H_R - H_L$. A direct computation shows that for ``large'' AdS-Schwarzschild black holes in d+1 dimensions ($r_H \gg \ell_{AdS}$), the dependence of black hole's entropy on temperature is:
 \be \label{ent-temp}
 S_B(T) \propto T^{d-1}.
 \ee
 The AdS/CFT duality implies that the black hole's partition function is equal to that of the boundary theory and, consequently, the entanglement entropy associated with one side equals the black hole's entropy in the bulk. 
 Entanglement, in general, has a non-local quantum nature; the action of a unitary operator on one side of an entangled state does not change the amount of entanglement between the two sides of the state.\\
 However, it is possible to change the states' entanglement entropy by coupling their two sides. The simplest example capturing this idea is the Bell pair:
 \be
 \frac{\ket{\uparrow \uparrow} + \ket{\downarrow \downarrow}}{\sqrt{2}}, 
 \ee
 with entanglement entropy equal to $\ln 2$. While the entanglement entropy remains unchanged under a local unitary operator, it is easy to construct a unitary operator which transforms the state to a pure state $\ket{\uparrow \uparrow}$.\\
 Similarly, one expects the entanglement entropy of the TFD to change under the unitary evolution $U(t) = e^{-iHt}$ where the Hamiltonian is 
\be \label{Hamilint}
H = H_0 + H_{int},~~~~~H_0=H_L+ H_R,~~~~~~
  H_{int} = \frac{g}{N} \sum_{i=1}^N \int ~ d^{d-1}x ~ \phi_L^i(\vec x) \phi_R^i(\vec x),
\ee
where $\phi_{L(R)}$ acts on the left (right) side of the black hole in the bulk. This Hamiltonian was used in \cite{Gao17} as a model for traversable wormholes; see also \cite{MSY17B,Mal18}. To compute the amount of entanglement change, we will prepare the $\TFD$ at $t=0$ and evolve it with U(t):
\be
\ket{\widetilde{\TFD}(t)} = U(t)~ \ket{\TFD}.
\ee
The entanglement entropy can be computed using the replica trick:
\be \label{replica}
S = -\p_s \ln \Tr \rho^s \Big|_{s=1}.
\ee
Direct computation yields that $\Delta S_{EE}$ to leading order in $g$ is:
\be \label{entanglement change}
\Delta S_{EE}(t) = ig \calS_{d-1} \int_0^t \, du~~~ \frac{d}{ds}  \bigg( G_{s\beta} \Big(2iu + \frac{\beta}{2} \Big) -G_{s\beta} \Big(-2iu + \frac{\beta}{2} \Big) \bigg)\bigg|_{s=1},
\ee 
where $G_{s\beta} \Big(2 i u + \frac{\beta}{2}\Big) = \bigg{\langle} \phi^i(u,\vec x) \phi^i(-u+i\frac{\beta}{2},\vec x)\bigg{\rangle}_{s\beta}$ is the two-point function at inverse temperature $s\beta$. Notice that the two-point function in \ref{entanglement change} is space independent, and so the integral over the spatial part gives the volume of the transverse direction, denoted by $\calS_{d-1}$. As a result, since the entropy of $\TFD$ is a function of its temperature, i.e. relation \ref{ent-temp}, one expects the temperature to change, and so does the black hole's size \footnote{For simplicity, we assume the black hole is parametrized by one parameter, its mass.}. On the other hand, when the system reaches the equilibrium, the new temperature associated with $\widetilde{\TFD}$ can be read from the two-point function whose computation needs  information about the theory's higher point functions. \\
This observation motivated us to study the dynamics of $\widetilde{\TFD}$ explicitly for a simple model, the Jackiw-Teitelboim (JT) gravity \cite{Teit83, Jackiw84}.
The JT gravity appears as the near-horizon limit of the four-dimensional charged black holes. The bulk is fixed to be AdS$_2$, and the dynamical degrees of freedom correspond to the reparametrization of the boundary whose dynamics are given by the Schwarzian action. The action initially appeared as the low energy limit of the SYK model \cite{Kisuh18A, MS16, SaYe93}. Computing the quantity \ref{entanglement change} in this model yields \ref{deltaent}:
\be \label{deltent1}
\Delta S_{EE} (t) = \frac{\pi b g}{2J} \Big(\frac{\pi}{\beta J}\Big)^{2\Delta-1} \bigg( 1 - \frac{1}{\Big(\cosh (\frac{2\pi t }{\beta})\Big)^{2\Delta}} \bigg),
\ee
which, after $t \sim \frac{2\pi}{\beta}$, implies the new temperature (\ref{temp}):
\be \label{temp1}
\tilde \beta_{EE} = \beta \Bigg( 1 - \frac{\pi b g}{2 J S} \Big(\frac{\pi}{\beta J} \Big)^{2\Delta-1} \Bigg).
\ee
On the other hand, the temperature of $\widetilde\TFD$, $\tilde \beta$, can be read from the two-point function after reaching the equilibrium. In computing the two-point function to leading order in the coupling, the four-point function of the theory is needed. A priori, these two temperatures do not have to be equal. In fact, they always satisfy $\beta \le \beta_{EE}$, as a local perturbation to one side of the $\TFD$ always raises its temperature while $\beta_{EE}$ remains unchanged. Equality of the two temperature means that the interaction \ref{Hamilint} acts \emph{coherently} on $\TFD$.  Note that the relations \ref{deltent1} and~\ref{temp1} are general and independent of the types of the fields.  the contributing modes to the four-point function are also generic and couple to all the fields, which in our case are the Schwarzian modes. Such modes are, indeed, the Goldstone modes associated with the spontaneous breaking of the reparametrization symmetry. In fact, the computation of the two-point function in section \ref{thermalization} confirms $\beta = \beta_{EE}$. \\
Our microscopic computation of the entanglement change  \ref{deltent1} can be rederived from a coarse-grained quantity, the Casimir associated with the $SL(2,\mathbb R)$ symmetry of the Schwarzian, \ref{noether}. More precisely, the entanglement entropy, to leading order in the coupling, can also be computed from \cite{Kisuh18B}\footnote{I am grateful to Alexei Kitaev for pointing this out to me.}:
\be \label{CG}
S(u) = 2\pi \sqrt{Q_R (u)},
\ee
which is computed over the solution to the equation of motion \ref{Energyconserv}. The exact match between the two quantities to the second order in the coupling \ref{entropy2} and \ref{entropy12} may confirm that \ref{CG} renders the  entanglement entropy dynamics after the quench by the interaction Hamiltonian \ref{Hamilint}.  While our computation of the coarse-grained entropy is for a one-dimensional theory, \ref{entanglement change} is true at any dimensions. The rest of the paper is organized as follows:  in section \ref{sec:two}, we will give a brief review of the JT gravity and the Schwarzian theory. In section \ref{sec:3}, we consider the $\widetilde \TFD$ and compute its entanglement entropy for a one dimensional large N field theory. As was pointed out earlier, to compute such a quantity, we only need to know the two-point function of the theory. In \ref{secondorderentang}, we  compute the second-order correction to the entanglement entropy. For that, we assume that we are at strong coupling where at low energy limit the theory is effectively described by the Schwarzian action, and the four-point function is given by \ref{schfourpt}.  Our computation using the replica method leads to $\ref{entropy2}$. In \ref{thermalization}, we compute the temperature of $\widetilde \TFD$ from the two-point function of two probing fields inserted in one side of the state. As was discussed earlier, this temperature is associated with the coarse-grained entropy of $\widetilde \TFD$, which is equal to the temperature associated with the entanglement entropy. In \ref{EE from e.q.m}, we will write an effective action for the system, which is two copies of the Schwarzian action, one for each side of the thermofield-double with the interaction term proportional to the two-point function of the microscopic fields. One can solve the equation of motion with the thermofield-double as the initial condition and find the position of the physical boundary. Moreover, such an action has the $SL(2,R)$ symmetry. Amazingly, the value of the Casimir function associated with the corresponding Noether charges gives the entanglement entropy that we computed microscopically. 
\section{Preliminaries}
\subsection{A brief review of the JT gravity and the Schwarzian theory} \label{sec:two}
The Jackiw-Teitelboim (JT) gravity \cite{Teit83,Jackiw84,Almheiri14} describes the near-horizon limit of a near extremal charged black hole in four dimensions, which can be derived from the spherical reduction of the Reissner-Nordstrom solution \cite{MBHE}. It is given by the following action:
\be \label{effaction}
 \begin{aligned} &
 S = \frac{\phi_s}{16\pi G} \bigg[ \int \, d^2x \sqrt{-g} \, R ~~+ 2\int \, \calK d\ell \bigg] +~~ \frac{1}{16\pi G} \bigg[ \, \int ~d^2 x \sqrt{-g} \Big( R +2 \Big) \phi +2\int \, \phi_b \calK d\ell \bigg],
 \end{aligned}
 \ee
where $\phi_s$ is the zero temperature entropy of the four-dimensional black hole. We also added the Hawking-Gibbons term to have a consistent variation.
  The first term is entirely topological, and the second bracket renders the dynamics. In the absence of the matter field, the dynamical action in the Euclidean time can be written as:
 \be \label{action}
 S = -\frac{1}{16\pi G} \bigg[ \, \int ~d^2 x \sqrt{g} \Big( R +2 \Big) \phi +2\int \, \phi_b~ (\calK-1) d\ell \bigg],
\ee
where the counterterm $\frac{1}{8\pi G}\int \, \phi_b~ d\ell $ was added to the action to make it finite. Variation with respect to the dilaton field yields:
 \be
 R+2 = 0,
 \ee
 \begin{figure}
 \centering
 \includegraphics[scale=.3]{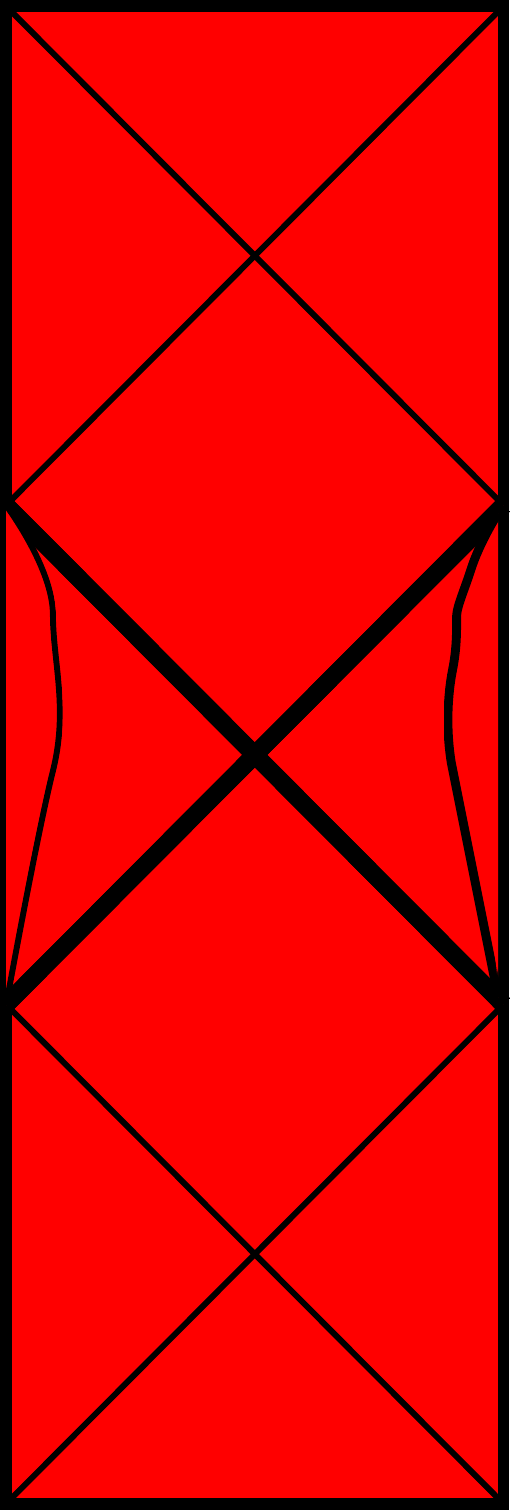}
 \caption{Here we assume that there is a sharp cutoff that separates the AdS region from the rest of the spacetime. This boundary can have arbitrary fluctuations }
 \label{fig:Bfluct}
 \end{figure}
 while the variation with respect to the metric gives the equation of motion for the dilaton field:
 \be \label{deom}
 \nabla_\mu \nabla_\nu \phi - g_{\mu\nu} \nabla^2 \phi + \phi g_{\mu\nu} = 8\pi G ~ T_{\mu\nu}.
 \ee
  Intergating out the dilaton, only the Hawking-Gibbons term remains:
 \be \label{eff}
 -\frac{1}{8\pi G} \, \int ~ \phi_b ~ (\calK-1) d\ell.
 \ee
For the case of the zero temperature which corresponds to the Poincare half plane,
 \be
 ds^2 = \frac{dt^2+ dz^2}{z^2}.
 \ee
 The boundary of this space is located at $z=0$, and as we approach it, the affine length will grow as $\frac{1}{z}$. Therefore, it is convenient to define the regularization parameter $\vep$ and define the ``physical boundary'' to be parametrized by the affine parameter $u$ so that 
 \be \label{affine}
 \frac{t'^2+z'^2}{z^2} = \frac{1}{\vep^2},
 \ee
 where derivative is with respect to $u$. To leading order in $\vep$,  \ref{affine} implies that the equation for the physical boundary will take the form:
 \be \label{bdy}
 \Big(t(u), z(u) \Big) = \Big( t(u), \vep t'(u) \Big).
\ee 
Computing the extrinsic curvature of the physical boundary yields the Schwarzian:
\be
\calK -1 = \vep^2 \Sch(t,u).
\ee
On the other hand, $\phi$ is a field with dimension two, and close to the boundary, it behaves as $\phi_b = \frac{\phi_r(u)}{\vep}$. We also have $d\ell = \frac{du}{\epsilon}$. We define the boundary as a curve on which the value  of the dilaton is  the constant $\phi_r$. Plugging into \ref{eff}, we will get the regularized action for the Poincare patch \cite{MSY17A}:
 \be \label{Sch}
 S = \frac{-\phi_r}{8\pi G} \, \int \, du \, \Sch(t(u), u),
 \ee
where the integrand is the Schwarzian derivative $\Sch(t(u), u) = \left(\frac{t''}{t'} \right)' - \frac{1}{2} \left( \frac{t''}{t'} \right)^2$ . The equation of motion together with \ref{bdy} determines the location of the boundary. One can think of the variable $u$ as the physical time and $t(u)$ as its arbitrary reparametrization. In the absence of the matter field in the bulk one can solve \ref{deom}:
 \be \label{deom}
 \phi(t,z) = \frac{\alpha(z^2+t^2) + \beta t + \gamma}{z}.
 \ee
On the boundary, it will take the form:
\be \label{beom}
\phi_r = \frac{\alpha t^2(u)+\beta t(u) + \gamma}{t'(u)}.
\ee
The solution associated to zero temperature is $t(u) = u$, $\phi(t,z) = \frac{\phi_r}{z}$. The finite temperature solution is associated with compactifying the time coordinate,
\be \label{compact}
t = \tan \frac{\varphi (u)}{2}.
\ee
Under this transformation, the action \ref{Sch} becomes:
\be
S = \frac{-\phi_r}{8\pi G} \int  du \, \left(\Sch(\varphi(u), u) + \frac{1}{2} \varphi'^2 \right).
\ee
We are interested in the solution
\be \label{th-solution}
\varphi(u) = \frac{2\pi u }{\beta},
\ee
with the free energy given by:
\be
 F = -\frac{\pi \phi_r}{4 G} ~T^2,
\ee
and other thermodynamic quantities equal to
\be
S = S_0 +\frac{\pi \phi_r}{2 G} ~T,~~~~~~~~~~E = \frac{\pi \phi_r}{4 G} ~T^2.
\ee
Here, $S_0$ comes from the topological term in \ref{effaction}. To understand the geometry of such a configuration, we will go to the global coordinate. In the Lorentzian signature, the boundary coordinate times are related by:
\be
t = \tanh \frac{\pi u}{\beta} = \tan \frac{\eta_R}{2} ,~~~~ -\frac{\pi}{2} \le \eta_R \le \frac{\pi}{2}.
\ee
\begin{figure}
\centering
\includegraphics[scale=.3]{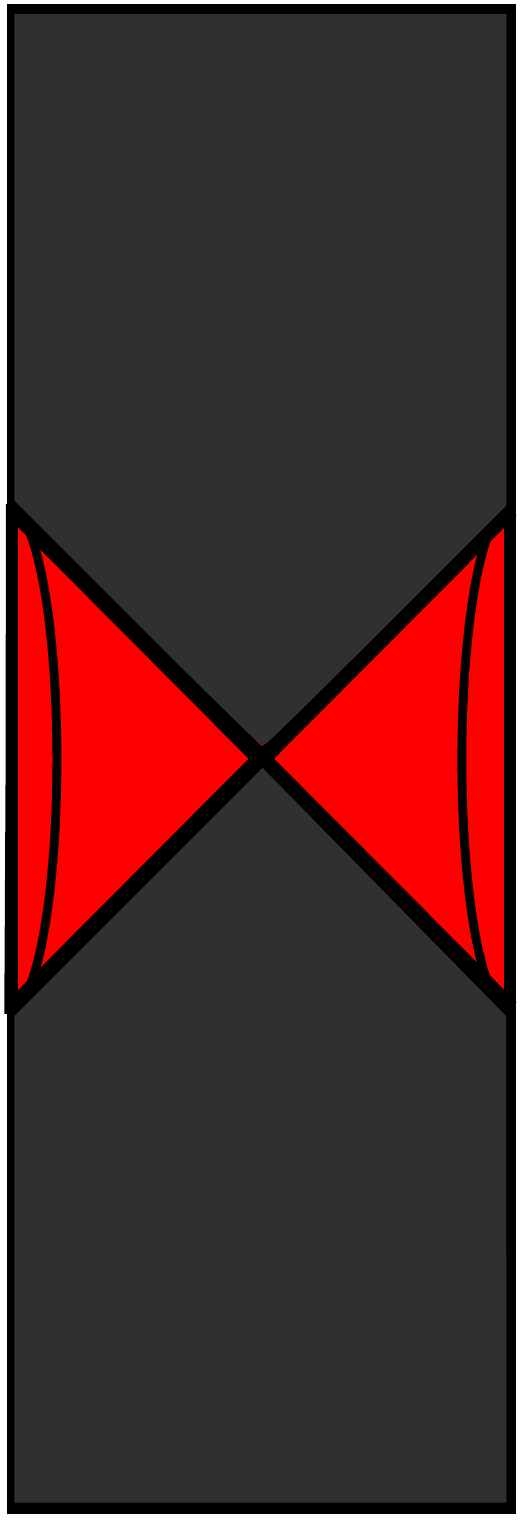}
\caption{The embedding of the two-dimensional black hole in global AdS$_2$ }
\label{fig:BH}
\end{figure}
In global coordinates, the thermofield-double will be represented by 
\be \label{TFD}
\Big(\eta \,\,, \, \sigma_r\Big)= \Big(2 \arctan \tanh \frac{\pi u}{\beta} \, , \, \frac{\pi}{2}- \frac{2\pi \epsilon}{\beta} \frac{1}{\cosh \frac{2 \pi u}{\beta}}\Big).
\ee
Defining $\eta' = e^\phi$  with the  lagrange multiplier $P_\eta$, the action will transform to
\be
\begin{aligned}
&- \int Sch \Big(\tan \frac{\eta}{2},\tilde u \Big) \, d \tilde u =  S =  \,\, \int \, d \tilde u \, \bigg( \frac{1}{2} (\phi'^2 - e^{2\phi}) +P_\eta(\eta' - e^\phi) \bigg) = \int \, du \, \bigg[ P_\eta \eta'+P_\phi \phi' -H \bigg] \\&
H =  \frac{P^2_\phi}{2}+\frac{1}{2} e^{2\phi}+P_\eta e^\phi, ~~~~~ \tilde u \equiv \frac{ 8\pi G \, u}{ \phi_r}.
\end{aligned}
\ee
Here derivative is with respect to $\tilde u$. Also  $(\phi, P_\phi)$ and $(\eta, P_\eta)$ are conjugate variables, and H is the Hamiltonian. The equation of motion is given by:
\be
\begin{aligned} \label{solution1}
&P'_\phi = -( e^{2\phi}+P_\eta e^\phi), ~~~~~\phi' = P_\phi, \\
&\eta' = e^\phi, ~~~~~~~~~ P'_\eta = 0.
\end{aligned}
\ee
The Schwarzian action also has the $SL(2,R)$ symmetry. The conserved charges are:
\be
\begin{aligned} \label{noether}
&Q_1 = \cos \eta (P_\eta+e^\phi)- \sin \eta P_\phi,\\
&Q_2 = \sin \eta (P_\eta+e^\phi)+\cos \eta P_\phi, \\
&Q_3 = P_\eta.
\end{aligned}
\ee
They satisfy $\{Q,H\} =0$, where the Poisson bracket is defined with respect to the conjugate variables $(\phi, P_\phi)$ and $(\eta, P_\eta)$. 

\subsection{Contribution of the Schwarzian modes to the four-point function}

In this section, we assume that we have a large N field theory, bosonic or fermionic,  whose low energy limit is described by the Schwarzian action. More precisely, we assume that the action has the form:
\be
S = S_1 - N \alpha \int_0^{2\pi} ~ d\theta ~ Sch \Big( e^{i\vp(\theta)} , \theta \Big), ~~~~~ \alpha = \frac{\phi_r}{4 N \beta G},
\ee
where $S_1$ is conformal and has the reparametrization symmetry which fixes the two-point function of operators of dimension $\Delta$.  An example of such a theory is the SYK model \cite{Kisuh18A, MS16, SaYe93}.
In the rest of the section, we compute the leading contribution of the Schwarzian action to the four-point function. First, we have
\be
\begin{aligned} &
\tilde G(\theta_1 , \theta_2)= \blangle \phi^i(\theta_1) \phi^i (\theta_2) \brangle = b ~\frac{ \vp'^{\Delta}(\theta_1)\vp'^{\Delta}(\theta_2)}{\Big( \sin \frac{\vp(\theta_1) - \vp(\theta_2)}{2} \Big)^{2\Delta}} =  G(\theta_1,\theta_2) \Big( 1+ \frac{\delta G}{G} \Big), \\&
G(\theta_1, \theta_2) = b \Big( \sin \frac{\theta_1 - \theta_2}{2} \Big)^{-2\Delta},~~~~ \frac{\delta G}{G} = \Delta \Big( \delta \vp'(\theta_1) + \delta \vp'(\theta_2) - \frac{\delta \vp (\theta_1) - \delta \vp (\theta_2)}{\tan \frac{\theta_1 - \theta_2}{2}} \Big).
\end{aligned}
\ee
Consider the fields $\phi_1$ with dimension $\Delta_1$ and $\phi_2$ with dimension $\Delta_2$.  Their four-point function has the following form:
\be
\begin{aligned} &
\frac{1}{N^2} \sum_{i,j}~\blangle \phi_1^i (\theta_1) \phi_1^i (\theta_2) \phi_2^j (\theta_3) \phi_2^j (\theta_4) \brangle = \\&   G_1(\theta_1-\theta_2)~ G_2(\theta_3-\theta_4) \begin{cases} \Big( 1 + \calF ^{TO}(\theta_1,\theta_2, \theta_3, \theta_4) \Big)~~~~~~0\le \theta_1 \le \theta_2 \le \theta_3 \le \theta_4 < 2\pi \\
\pm\Big( 1+ \calF^{OTO} (\theta_1,\theta_2, \theta_3, \theta_4) \Big)~~0\le \theta_1 \le \theta_3 \le \theta_2 \le \theta_4 < 2\pi 
\end{cases},
\end{aligned}
\ee
where $\pm$ is for bosons and fermions, respectively, and $\calF$ denotes the connected part of the four-point function and it is equal to:
\be \label{4point1}
\begin{aligned}
\calF (\theta_1,\theta_2, \theta_3, \theta_4) = \blangle \frac{\delta G_1(\theta_1, \theta_2)}{G_1}~ \frac{\delta G_2 (\theta_3, \theta_4)}{G_2} \brangle_c.
\end{aligned}
\ee
The r.h.s can be computed by expanding the Schwarzian action around \ref{th-solution} to second order in the soft modes and computing the associated two-point function. The final answer is\cite{Kisuh18A, MS16}:
\be \label{schfourpt}
\calF \Big(\theta_1,\theta_2,\theta_3,\theta_4\Big) = \frac{4\Delta_1\Delta_2}{ S}\begin{cases}
\Bigg( 1 - \frac{\theta}{2~\tan \frac{\theta}{2}} \Bigg)\Bigg( 1 - \frac{\theta'}{2~\tan \frac{\theta'}{2}} \Bigg)~~~~~~~~~~~~~~~~~~~~~~~~~~~~~~~~~~~~~~~~~~~(TO) \\
\frac{-\pi \sin \Delta \theta_+}{2 \sin \frac{\theta}{2} ~\sin \frac{\theta'}{2}} - \frac{\pi ( \pi - 2\Delta \theta_+ )}{4 \tan \frac{\theta}{2} \tan \frac{\theta'}{2}} +\Bigg( 1 +\frac{\pi - \theta}{2~\tan \frac{\theta}{2}} \Bigg)\Bigg( 1 + \frac{\pi - \theta'}{2~\tan \frac{\theta'}{2}} \Bigg) ~~~~~~~~~(OTO)
\end{cases}.
\ee
\section{Entanglement entropy of $\widetilde{\TFD}$} \label{sec:3}
In this section, we will study the states that are produced by evolving the thermofield-double state $\ket{\TFD} \in \calH^* \otimes \calH \equiv \calH_L \otimes \calH_R$ by double-traced operators coupling both sides of the thermofield-double:
\be
\ket{\widetilde{\TFD}(t)} = \widetilde U(t) ~\ket{\TFD}
\ee
where the unitary operator $U(t)$ has the form
\be \label{unitary1}
\begin{aligned} &
\widetilde U(t) = ~\TT \exp \Big( -i \int_0^t du~\widetilde H (u) \Big) \\&
\widetilde H(u) = H_L+H_R+ H_{int}(u),~~~~~~~
H_{int}(u) = \frac{g(u)}{N} \sum_{i=1}^{N}\phi^i_L \phi^i_R~~~~~~N \gg 1,
\end{aligned}
\ee
and $H_L = H^* \otimes 1$ and $H_R = 1 \otimes H$. Note also that the time dependence of $\widetilde H(u)$ is only through $g(u)$.
Here we assume that the fields $\phi^i$s are bosonic  with dimension $\Delta$\footnote{ One can also take fermionic fields. In this case, the terms $\phi^j_L \phi^j_R$ in the Hamiltonian should be modified to $i\phi^j_L \phi^j_R$ for the Hamiltonian to be Hermitian. However, the results remain unchanged.}. Note that the time direction on the left side is the opposite to the one on the right side. In other words, the generator of time evolution is $H = H_R - H_L$ \footnote{$\phi_R(t) = e^{iH_R t} \phi_R e^{-iH_R t}$, while   $\phi_L(t) = e^{-iH_L t} \phi_L e^{iH_L t}$.}. This means the Hamiltonian $H=H_L + H_R$, while it takes the right fields forward in time, it takes the left fields backward in time. Therefore, in the interaction picture, the evolution operator takes the following form:
\be \label{unitary2}
U_I(t) = \TT \exp \Big( -i \int_0^t du ~ H_{I} (u) \Big), ~~~H_I(u) = \frac{1}{N} \sum_{i=1}^{N} ~g(u)\phi^i_L(-u) \phi^i_R(u).
\ee
We further assume that the unperturbed theory, $(H_{int} = 0)$, at low temperature has an approximate conformal symmetry, i.e. the two point function is conformal and in Euclidean time it is: \footnote{Here,  $J$  has dimension of energy, and we extracted it from numerator to make both numerator and denominator manifestly dimensionless.}
\be \label{2pt}
\langle \phi_R^i (\tau) \phi_R^j(\tau') \rangle = \frac{b \delta_{ij}}{\Big(\frac{\beta J}{\pi} \sin \frac{\pi (\tau-\tau')}{\beta}\Big)^{2\Delta}},
\ee
while higher point functions are described by the  Schwarzian modes. To construct the density matrix associated with the right side, we start with the thermofield-double which can be diagrammatically represented as:
\be
\ket{\TFD} ~~~~~~~~~~~~\leftrightarrow ~~~~~~~~~ \qquad \includegraphics[scale=0.3, valign=c]{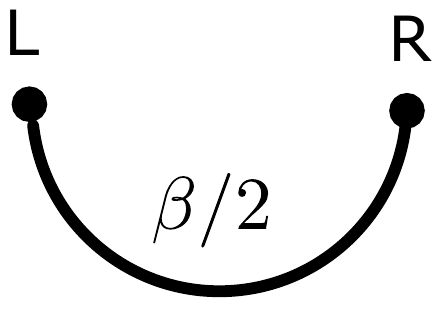}
\ee
 Then for $0 \le t_1,t_2 \le t$, we have the following diagrammatic representations: 
\be \label{rules}
 \bra{\TFD} \phi_L(-t_2) \phi_R(t_1)=~~ \figbox{0.3}{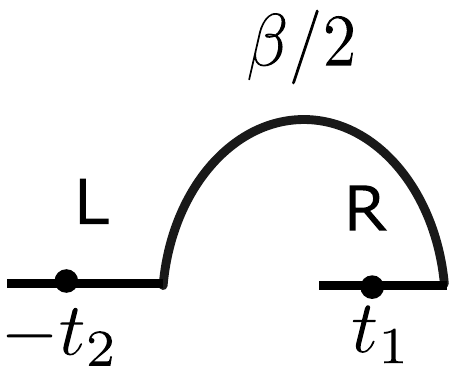} ~~~~~~~~~~~\phi_R(t_1) \phi_L(-t_2) \ket{\TFD} =~~ \figbox{0.3}{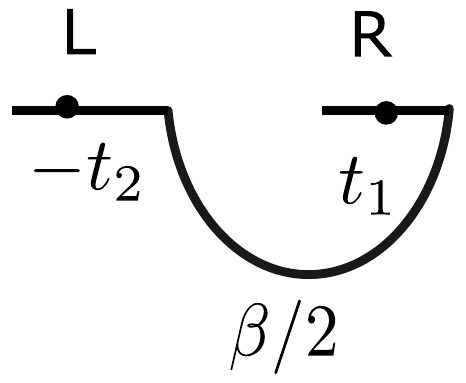}
\ee

Now, consider two copies of the thermofield-double with the insertions from the unitary operator \ref{unitary2}
\be \label{twoTFD}
 \widetilde U_I(t) \ket{\TFD} \bra{\TFD} \widetilde U_I^{-1}(t)  ~~~~~~\leftrightarrow~~~~~ \includegraphics[scale=0.4, valign=c]{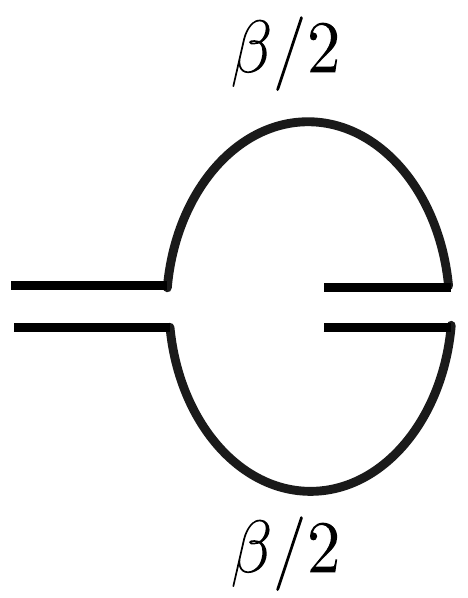}
\ee
Then we can use the identity 
\footnote{This is indeed a Euclidean rotation with angle $\pi$ by the modular operator. More abstractly, the operator algebras associated with the left and the right wedges, $\calA_R$ and $\calA_L$,  are commutants. According to the Tomita-Takesaki theory there is an anti-unitary operator $\calJ$ such that $\calJ \calA_L \calJ = \calA_R$. In our case $\calJ = \CRT$ where C, R, T  are charge conjugation, reflection and time reversal operators, respectively.  Moreover, we have $\calJ A^\dagger_R \ket{TFD} = \Delta^{\frac{1}{2}} A_R \ket{TFD}$, where $\Delta = e^{-\beta H}$. Note that the l.h.s and r.h.s belong to the left wedge. For $A_R$ to be  a scalar field we have the relation \ref{lr}. For fermionic fields, we have $\phi_L (t) \ket{\TFD} = i \phi_R (t+i\frac{\beta}{2}) \ket{\TFD}$. See \cite{BW76,Witten18,Haag92} for more details.}
\be \label{lr}
\phi_L (-t) \ket{\TFD} = \phi_R (-t+i\frac{\beta}{2}) \ket{\TFD} \
\ee
to transform all the left fields to the right ones and trace out the left side. This corresponds to gluing the left end points of the contour,
\be \label{rightdensity}
\rho_R(t) = \Tr_L \Big( \widetilde U(t) \ket{\TFD} \bra{\TFD} \widetilde U^{-1}(t) \Big) ~~~~~~~~~~~\leftrightarrow ~~~~~~~\includegraphics[scale=0.3, valign=c]{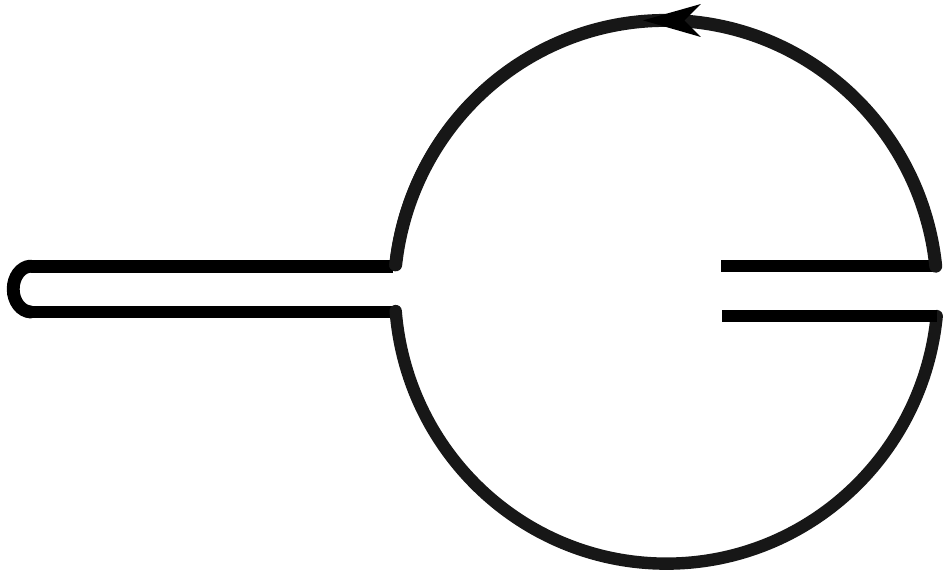}
\ee
  An explicit expression for $\rho_R(t)$ will take the following form:
\be \label{RDS}
\begin{aligned} 
 \rho_R(t) =& \sum_{n,m} \frac{(i)^m (-i)^n}{N^{n+m} n! \, m!}  \int_0^t (du_1 \cdots du_n)(du'_1 \cdots du'_m)g(u_1) \cdots g(u_n) g(u'_1) \cdots g(u'_m) \\
 & \sum_{i_1j_1,...,i_{n}j_{n}=1}^N   \frac{e^{-\beta H_R}}{Z(\beta)}   \bigg[ \phi^{i_1}(u_1-i\beta) \cdots \phi^{i_{n}}(u_n-i\beta) \phi^{i_n}(-u_n-i\frac{\beta}{2})\cdots
\phi^{i_1} (-u_1-i\frac{\beta}{2})  \\
&~~~~~~~~~~~~~~~~~~~~~~~~~~~~~~~~~~~~~\phi^{j_1} (-u'_1-i\frac{\beta}{2}) \cdots \phi^{j_m}(-u'_m-i\frac{\beta}{2})\phi^{j_m}(u'_m)\cdots \phi^{j_1}(u'_1)   \bigg], \\&~~~~~~~~~~~~~~~~~~~ (u_1 \ge \cdots \ge u_n \ge 0,~~u'_1 \ge \cdots \ge u'_n \ge 0).
\end{aligned}
\ee
Note that, in the above expansion associated with each operator there is a ``mirror operator''; an operator with the same index whose argument's real part is negative of the original operator and its imaginary part is shifted by $-i\frac{\beta}{2}$, e.g., $\phi^i(u)$ and $\phi^i(-u-i\frac{\beta}{2})$. Such operators are depicted by the same color in \ref{rightdensity}.
One can construct the Renyi entropy by gluing $s$ copies of $\rho_R$ and compute the entanglement entropy: 
\be
S_{EE} = \lim_{s \rightarrow 1} ~ \frac{1}{1-s} \log \Tr \rho_R^s.
\ee 
More explicitly, we consider $s$ copies of \ref{RDS}, which has the following schematic form:
\be
\Tr (\rho_R^s) = \Tr \Big( \frac{e^{-\beta H_R}}{Z(\beta)} \Big[ \cdots\Big]  \cdots \frac{e^{-\beta H_R}}{Z(\beta)} \Big[ \cdots\Big] \Big) =  \Tr \Big( \frac{e^{-s\beta H_R}}{Z^s(\beta)} \Big[\cdots\Big] \cdots \Big[\cdots\Big]\Big),
\ee
where inside each bracket there are fields and their mirrors as in \ref{RDS} and for simplicity, we dropped all the sums and integrals. Next, we moved all the $\frac{e^{-\beta H_R}}{Z(\beta)}$ to the left. When each of these terms is moved over a bracket, it will shift all the operators' arguments inside the bracket by $-i\beta$. This is not important since, at the end, we will break each bracket into the product of two-point functions, which are the function of time differences\footnote{This will also be the case when we compute the next order correction to the entanglement entropy where we need to use the four-point function  given in \ref{schfourpt}.}. As was just pointed out, the leading contribution to $\Tr (\rho_R^s)$ comes from  breaking each replica into a product of the two-point function of the fields with the same index at temperature $s\beta$. Doing so and considering the sum over indices in \ref{RDS}, we can exponentiate each bracket. Taking the logarithm of the result yields: 
\be
\begin{aligned} &
\ln \Tr \Big( \rho_R^s(t) \Big) = \ln \frac{Z_0(s\beta)}{Z_0^s(\beta)} - i s ~ \int_0^t ~g(u) du \Big( ~G_{s\beta}(2iu + \frac{\beta}{2}) - G_{s\beta}(-2iu + \frac{\beta}{2}) \Big). 
\end{aligned}
\ee
In the limit $s \rightarrow 1$, the first term in the right hand side is the entanglement entropy of $\ket{\TFD}$. Therefore, $\Delta S$ is:
\be
\begin{aligned} 
\Delta S &= i\Big(\frac{\pi}{\beta J} \Big)^{2\Delta} \int_{0}^u g(u) d u~~~~ \frac{d}{ds}\bigg[ \bigg(\frac{1}{\sin (\frac{2\pi i u }{s\beta}+\frac{\pi}{2s})} \bigg)^{2\Delta }-\bigg(\frac{1}{\sin (\frac{-2\pi i u }{s\beta}+\frac{\pi}{2s})}\bigg)^{2 \Delta} \bigg]_{s=1} \\
&= 2\pi\Delta~ b\Big(\frac{\pi}{\beta J} \Big)^{2\Delta} \int_{0}^u g(u) d u~~~~ \frac{\sinh \frac{2\pi u}{\beta}}{\Big(\cosh (\frac{2\pi u }{\beta})\Big)^{2\Delta+1}}.
\end{aligned}
\ee
\begin{figure}[t]
\centering
\includegraphics[scale=.65]{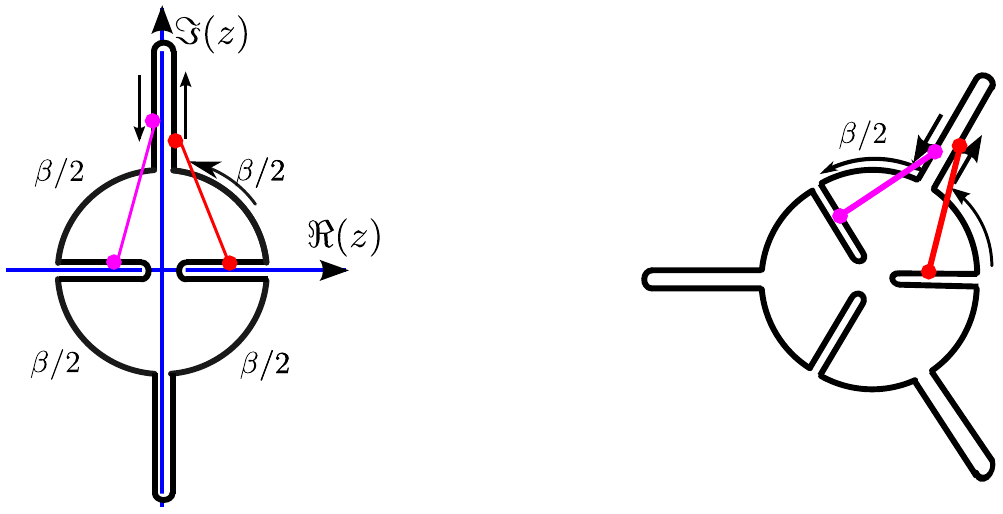}

\caption{The time contour for the second(left) and the third(right) Renyi entropy. The contours can be constructed by gluing two and three contours of \ref{rightdensity}. The insertions are represented by filled circles and the Wick's contractions are represented by solid lines. In the left figure, $z = e^{\frac{\pi (i \tau-t)}{\beta}}$, and in the right figure, $z=e^{\frac{2\pi (i \tau-t)}{3\beta}}$.}
\label{fig:Renyi23}
\end{figure}
Considering a quantum quench, $g(u) = g~ \delta (u)$, we can evaluate the integral:
\be \label{deltaent}
\Delta S_{EE} (t) = \frac{\pi b g}{2J} \Big(\frac{\pi}{\beta J}\Big)^{2\Delta-1} \bigg( 1 - \frac{1}{\Big(\cosh (\frac{2\pi t }{\beta})\Big)^{2\Delta}} \bigg).
\ee
As is clear,  $\Delta S_{EE}$ will saturate at~ $t \sim \frac{2\pi}{\beta}$ with the value:
\be \label{deltaentequ}
\Delta S^*_{EE} = \frac{\pi b g}{2J} \Big(\frac{\pi}{\beta J}\Big)^{2\Delta-1}.
\ee
One expects at this time the system to thermalize. Since $S \propto T$, the system thermalizes with the new temperature
\be \label{temp}
\tilde \beta = \beta ~ \bigg( 1 - \frac{\Delta S}{S} \bigg) = \beta ~\bigg( 1 - \frac{\pi b g}{2JS} \Big(\frac{\pi}{\beta J}\Big)^{2\Delta-1} + O(\frac{1}{S^2}) \bigg).
\ee
\subsection{Second order correction to the entanglement entropy \label{secondorderentang}}

The second-order correction comes from the connected part of the four-point functions in the OPE limit. There are two configurations of the fields. The first configuration corresponds to the pairs of fields with the same index that belongs to different replicas:
\be \label{nnest}
\begin{aligned} &
\frac{s(s-1)}{2} g^2 \int ~ du du' ~ G(2 i u + \frac{\beta}{2}) ~ G( 2i u' + \frac{\beta}{2}) \bigg[ \calF^{TO} (2iu + \frac{\beta}{2}, -2 i u' + \frac{\beta}{2}) +\calF^{TO} (2 i u' + \frac{\beta}{2}, -2 i u + \frac{\beta}{2}) \\&~~~~~~~~~~~~~~~~~~~~~~~~~~~~~~~~~~~~~~~~~~~~~~~~~~~~~~~~ -\calF^{TO} (2 i u' + \frac{\beta}{2}, 2 i u + \frac{\beta}{2}) -\calF^{TO} (-2 i u' + \frac{\beta}{2}, -2 i u + \frac{\beta}{2}) \bigg] \\&
= \frac{\pi^2 b^2 g^2}{2 S} \Big(\frac{\beta}{2\pi}\Big)^2 \Big(\frac{\pi}{\beta J}\Big)^{4\Delta} \Big(1 - \frac{1}{\cosh \frac{2\pi t}{\beta}} \Big)^2 (s-1).
\end{aligned}
\ee
\begin{figure}
\centering
\includegraphics[scale=.3]{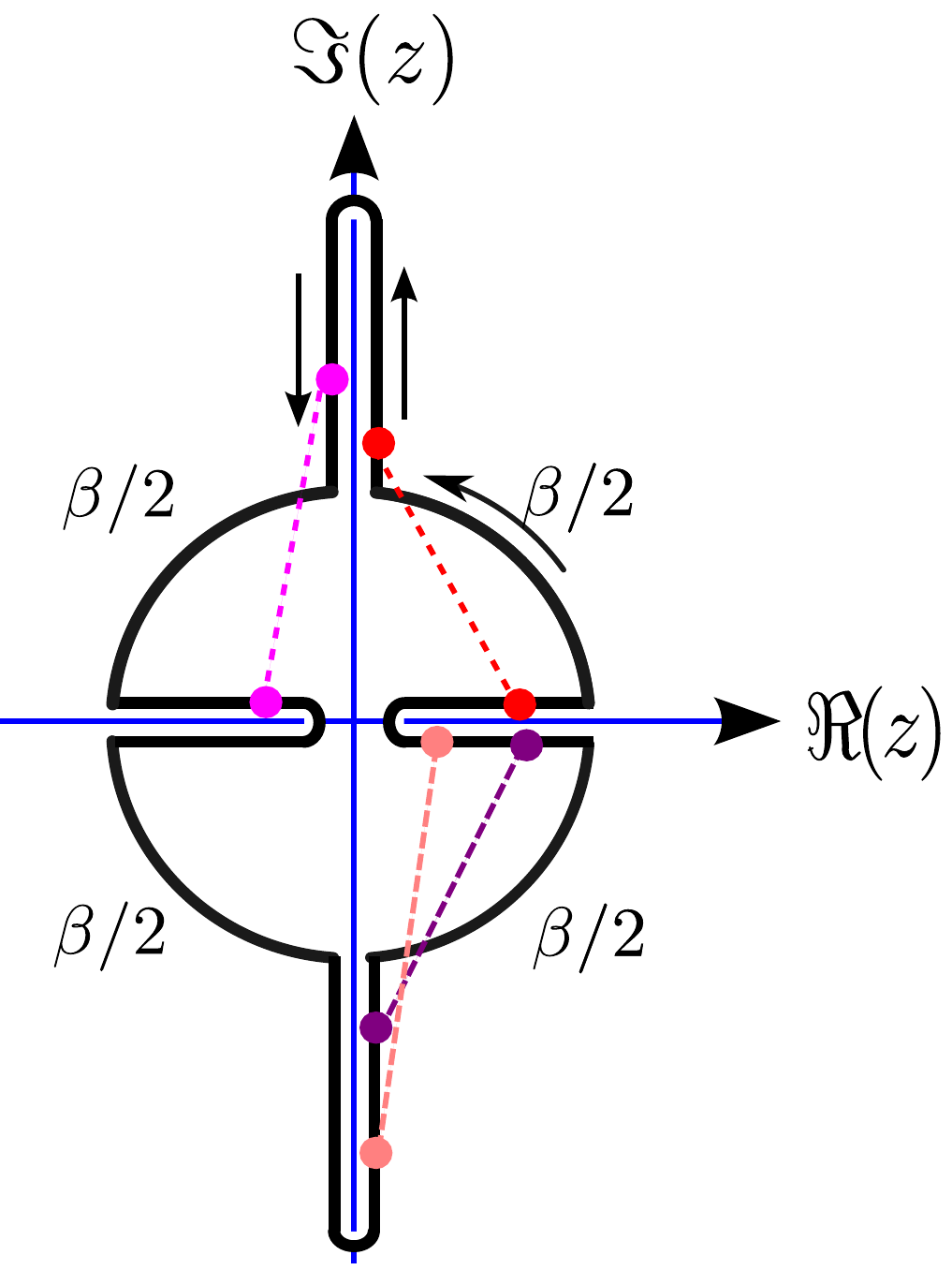}
\caption{The contribution of four-point functions in the second configuration that contributes to the entanglement entropy ($z = e^{\frac{\pi(i\tau-t)}{\beta}}$). There are two types: the first depicted by the pink and red dotted lines where the insertions are disjoint. The second type corresponds to the nested correlator.}
\label{figure:secondorder}
\end{figure} The second configuration consists of the pairs that belong to the same replica, but are located on different branches of the Keldysh contour and the pairs  which are located on the same branch of the Keldysh contour, which are nested pairs, see figure \ref{figure:secondorder}. The expression associated with the second configuration is equal to: 
\be \label{nest}
\begin{aligned} &
  s g^2  ~\int_0^t ~dudu' ~ G(-2 i u + \frac{\beta}{2}) ~ G(+ 2i u' + \frac{\beta}{2}) ~\calF (-2iu + \frac{\beta}{2}, 2 i u' + \frac{\beta}{2}) \\& -\frac{s}{2} g^2 \int_0^t ~ du~\int_0^t~du'~\bigg[ G(-2iu -\frac{\beta}{2} + s\beta) G( 2i u' + \frac{\beta}{2}) \calF_{\TT} (-2iu -\frac{\beta}{2} + s\beta, 2i u' + \frac{\beta}{2})+c.c. \bigg],
\end{aligned}
\ee
where the second line corresponds to $\blangle \TT \Big{\{} \phi^{i}(iu + \frac{\beta}{2}) \phi^j (iu' + \frac{\beta}{2}) \phi^j(-iu') \phi^i(-iu) \Big{\}} \brangle +\blangle \widetilde \TT \Big{\{} \phi^{i}(-iu + \frac{\beta}{2}) \phi^j (-iu' + \frac{\beta}{2}) \phi^j(iu') \phi^i(iu) \Big{\}} \brangle$. Both expressions are with respect to the inverse temperature  $s\beta$. Note that in the limit $s \rightarrow 1$, both \ref{nnest} and \ref{nest} vanish, which means we can compute their contribution to the entanglement entropy separately. Let us define $X = \frac{i u }{s} + \frac{\pi}{2s}$ and $Y =\frac{-i u }{s} + \frac{\pi}{2s}$, and assume $\beta = 2\pi$. We have $\blangle \TT \Big{\{} \phi^{i}(iu + \frac{\beta}{2}) \phi^j (iu' + \frac{\beta}{2}) \phi^j(-iu') \phi^i(-iu) \Big{\}} \brangle =$
\be
 \Big(\frac{\pi}{\beta J} \Big)^{4\Delta} ~ \frac{4 \Delta^2 b^2}{ S}
\begin{cases}
 \Big(1 - \frac{X}{\tan X} +\frac{\pi}{\tan X}\Big)\Big(1 - \frac{X'}{\tan X'} \Big) \frac{1}{\sin^{2\Delta} X ~\sin^{2\Delta} X'}~~~~~~~u>u' \\
 \Big(1 - \frac{X}{\tan X}\Big)\Big(1 - \frac{X'}{\tan X'}+\frac{\pi}{\tan X'}\Big)\frac{1}{\sin^{2\Delta} X ~\sin^{2\Delta} X'}~~~~~~~u'>u,
\end{cases}
\ee
and the anti-time-ordered four-point function is simply the complex conjugate of the above expression. So the expression for \ref{nest} will take the following form:
\be \label{secondorder}
\begin{aligned} &
-\frac{s g^2}{2} \frac{4 b^2\Delta^2}{S} \Big(\frac{\pi}{\beta J} \Big)^{4\Delta}\Bigg[\bigg[\int_0^t du \bigg( \frac{1}{\sin^{2\Delta} X} \Big(1 - \frac{X}{\tan X}\Big) - \frac{1}{\sin^{2\Delta} Y} \Big(1 - \frac{Y}{\tan Y}\Big) \bigg) \bigg]^2  \\& ~~~~~~~~~~~~~~~~~~~~~~~~~~~~~~~~~~~+ \bigg[\int_0^t du ~ \frac{1}{\sin^{2\Delta} X} \frac{\pi}{\tan X} \int_0^{u} ~ du' \frac{1}{\sin^{2\Delta} X'} \Big(1 - \frac{X'}{\tan X'}\Big) \\&
~~~~~~~~~~~~~~~~~~~~~~~~~~~~~~~~~~~+\int_0^t du ~ \frac{1}{\sin^{2\Delta} X} \Big(1 - \frac{X}{\tan X}\Big) \int_u^{t} ~ du' \frac{1}{\sin^{2\Delta} X'} \frac{\pi}{\tan X'} +c.c.\bigg]\Bigg].
\end{aligned}
\ee
We can use by-parts and some manipulations to simplify the second and the third line further: 
\be
\begin{aligned} &
-\frac{s g^2}{2} \frac{4 b^2\Delta^2}{S} \Big(\frac{\pi}{\beta J} \Big)^{4\Delta}\Bigg[\bigg[\int_0^t du \bigg( \frac{1}{\sin^{2\Delta} X} \Big(1 - \frac{X}{\tan X}\Big) - \frac{1}{\sin^{2\Delta} Y} \Big(1 - \frac{Y}{\tan Y}\Big) \bigg) \bigg]^2  \\&
~~~~~~~~~~~~~~~~~~~~~~~~~+\bigg[\frac{\pi^2 s}{8\Delta^2 \sin^{4\Delta}\frac{\pi}{2s}} - \frac{\pi^2 s}{4\Delta^2 \sin^{2\Delta}\frac{\pi}{2s}} \sin^{-2\Delta} X + \frac{\pi s^2}{4\Delta^2} X \sin^{-4\Delta} X + c.c. \bigg]\\&~~~~~~~~~~~~~~~~~~~~~~~~~-\frac{i\pi s}{2\Delta} (2-\frac{1}{2\Delta}) \Big( \int_0^t \frac{1}{\sin^{4\Delta} X} - c.c. \Big) +\frac{i\pi s}{\Delta} \Big(1-\frac{1}{2\Delta}\Big)\Big( \frac{1}{\sin^{2\Delta} X} \int_0^t \frac{1}{\sin^{2\Delta} X} - c.c.\Bigg].
\end{aligned}
\ee
Expanding around $s=1$ yields:
\be
\begin{aligned} &
\frac{-\pi^2 b^2 g^2}{2S} \Big(\frac{\beta}{2\pi} \Big)^2 \Big(\frac{\pi}{\beta J}\Big)^{4\Delta} \Bigg[4\Delta \frac{t\tanh t}{\cosh^{2\Delta }t} - 4\Delta \frac{1}{\cosh^{2\Delta} t} + \frac{4\Delta}{\cosh^{4\Delta} t} +4\Delta (2\Delta-1) \frac{\sinh t}{\cosh^{1+2\Delta} t} \int_0^t \frac{du}{\cosh^{2\Delta} u} \Bigg].
\end{aligned}
\ee
 Therefore, retrieving the $\beta$ dependence, the second order correction to the entanglement entropy, including \ref{nnest}, is
\be \label{entropy2}
\begin{aligned} &
\Delta S_{EE} = \frac{\pi^2 b^2 g^2}{2S} \Big(\frac{\beta}{2\pi} \Big)^2 \Big(\frac{\pi}{\beta J}\Big)^{4\Delta} \Bigg[ -\Big(1 - \frac{1}{\cosh^{2\Delta} \frac{2\pi t}{\beta}} \Big)^2 + \frac{4\Delta}{\cosh^{2\Delta} \frac{2\pi t}{\beta}} \bigg[ \frac{2\pi t}{\beta}\tanh \frac{2\pi t}{\beta} -1 + \frac{1}{\cosh^{2\Delta} \frac{2\pi t}{\beta}} \\&
~~~~~~~~~~~~~~~~~~~~~~~~~~~~~~~~~~~~~~~~~~~~~~~~~~~~~~~~~~~~~~~~~~~~~~~~~~~~~~~~+(2\Delta-1) \tanh \frac{2\pi t}{\beta} \int_0^\frac{2\pi t}{\beta} \frac{du}{\cosh^{2\Delta} u} \bigg] \Bigg].
\end{aligned}
\ee

\subsection{Thermalization \label{thermalization}}
\begin{figure}
\centering
\includegraphics[scale=.4]{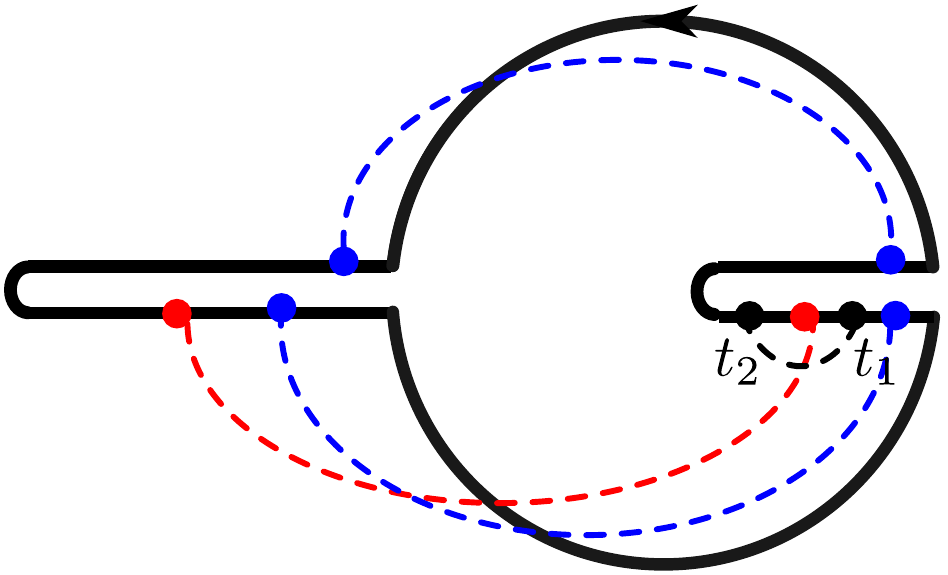}
\caption{The time contour for the four-point function using \ref{rules}. Here, we have $-\frac{\beta}{2} \le \tau \le \frac{\beta}{2}$. The figure displays different configurations in the computation of the two-point function. The probing fields are depicted by the black filled circles. The blue/red filled circles represent the field insertions from the unitary evolution. The red insertions corresponding to the middle unitary operator in \ref{tprb} make the out-of-time-ordered correlator with the probing fields.}
\label{fig:thermal}
\end{figure}

In this section, we will compute the temperature of the deformed thermofield-double state directly by studying the two point function of two probing fields inserted in the right side. In general,  for a system out of equilibrium, the two-point function $G(t_1,t_2)$ also depends on $\frac{t_1+t_2}{2}$.  In our case, the correction to the two-point function is of two types. The first type is the case where the interaction Hamiltonian makes a time-ordered correlation function with the two probing fields in the two point function, while the second type is the case where they make an out-of-time-ordered correlator; see Figure \ref{fig:thermal}. We expect the $\frac{t_1+t_2}{2}$ dependence of the two-point function to come from the second type, from the exponentially growing term in the OTOC configuration \cite{Shenker14}. However, as we will see the non-equilibrium part will be suppressed at the time of order $t \sim \frac{\beta}{2\Delta}$ due to exponential decay in the strength of the interaction Hamiltonian\footnote{More precisely, last line of \ref{twopointtherm} will be suppressed for $t_1,t_2 \gtrsim \beta$.}, and the system will equilibrate with the new temperature  much earlier than the scrambling time. 
 We insert the two probing fields of dimension $\Delta_1$ at time $t_1$ and $t_2$:
\be \label{tprb}
\frac{1}{N} \sum_{i=1}^N \bigg\langle{TFD}\bigg| \color{blue} U_I(t_0,t_2) \,\color{black} \phi^i(t_2) \,\color{red} U_I(t_2,t_1) \,\color{black} \phi^i(t_1)\color {blue} U_I(t_1,t_0) \color{black} \bigg|{TFD} \bigg\rangle,
\ee
Here, $t_0=0$ is the time TFD is prepared. We expand the three evolution operators inserted in the two-point function which to leading order in g yields:
\be \label{twopointtherm}
\begin{aligned} &
 G_{\beta} (t_2-t_1)\Bigg(1-i  \,\,  \bigg[ \int_{t_0}^{t_1} du \, g(u) \, \mathcal{F}^{TO}\bigg(i(t_2-t_1),2iu+\frac{\beta}{2}\bigg)\,\, G(2iu+\frac{\beta}{2}) \\&
 ~~~~~~~~~~~~~~~~~~~~~~~~~~~~~~~-\int_{t_0}^{t_2} du \, g(u) \, \mathcal{F}^{TO}\bigg(-2iu+\frac{\beta}{2},i(t_2-t_1)\bigg) \,\, G(2iu+\frac{\beta}{2})  \\&
~~~~~~~~~~~~~~~~~~~~~~~~~~~~~~~~+\int_{t_1}^{t_2} du \, g(u) \, \mathcal{F}^{OTO}\bigg(it_2, it_1,iu,-iu-\frac{\beta}{2}\bigg) G(2iu+\frac{\beta}{2})  \bigg]\Bigg).
\end{aligned}
\ee
For $t_1,t_2 > t_0$ after manipulation, we get:
\be
\begin{aligned} &
G((t_2-t_1)) \Bigg[1+\frac{\beta b g \Delta_1}{2 S (\frac{\beta J}{\pi})^{2\Delta}} \Bigg( \Big( 2 - \frac{1}{\cosh^{2\Delta} \frac{2\pi t_1}{\beta}} - \frac{1}{\cosh^{2\Delta} \frac{2\pi t_2}{\beta}} \Big) \Big(1 - \frac{\frac{\pi(t_2-t_1)}{\beta}}{\tanh \frac{\pi(t_2-t_1)}{\beta}} \Big) \\&
+\int_{t_1}^{t_2} ~ du~ \bigg[ \frac{\frac{\pi}{4} \cosh \frac{\pi (t_1+t_2)}{\beta}}{\sinh \frac{\pi(t_2-t_1)}{\beta}} \frac{1}{\cosh^{2\Delta+1} \frac{2\pi u}{\beta}} \bigg] - \frac{\pi}{2} \coth \frac{\pi (t_2-t_1)}{\beta} \int_{t_1}^{t_2} ~du~ \frac{1}{(\cosh \frac{2\pi u}{\beta})^{2\Delta}} \Big( 1 - \frac{\frac{2\pi u}{\beta}}{\coth \frac{2\pi u}{\beta}} \Big) \\&
+\bigg[\frac{\frac{\pi (t_1+t_2)}{8\Delta}}{\tanh \frac{\pi (t_2-t_1)}{2}} \Big( \frac{1}{\cosh^{2\Delta}(\frac{2\pi t_2}{\beta})} - \frac{1}{\cosh^{2\Delta} (\frac{2\pi t_1}{\beta})} \Big)\Bigg) \Bigg].
\end{aligned}
\ee
 At $t_1,t_2 \sim \beta$, one can approximate $\cosh \frac{2\pi t}{\beta} \sim \frac{e^{\frac{2\pi t}{\beta}}}{2}$ and see that all these terms are negligible. Therefore, the terms that survive are:
 \be
 G(t_2-t_1) \bigg[1+\frac{\beta b g \Delta_1}{ S (\frac{\beta J}{\pi})^{2\Delta}} \Big(1 - \frac{\frac{\pi(t_2-t_1)}{\beta}}{\tanh \frac{\pi(t_2-t_1)}{\beta}} \Big) \bigg] = G_{\tilde \beta} (t_2-t_1).
 \ee
 where 
 \be \label{cgtemp}
 \tilde \beta = \beta \Big( 1 - \frac{\pi b g}{2J S}(\frac{\pi}{\beta J})^{2\Delta-1} \Big).
 \ee
 This matches the temperature that was predicted from computing the entanglement entropy, i.e. \ref{temp}.

\subsection{The entanglement entropy from the equation of motion} \label{EE from e.q.m}
In this section, we will derive the entanglement entropy from the equation of motion for the boundary. For that, we will use the global AdS coordinate. The relation between coordinate times in Poincare, Schwarzschild, and global time coordinates is:
\be
t = \tanh \frac{\pi \tau}{\beta} = \tan \frac{\eta}{2},
\ee
and so the two-point function between the left and the right boundary in terms the global AdS coordinate is given by:
\be
\blangle \phi_R^i(u_1) \phi_L^j(-u_2) \brangle = b ~ \left( \frac{\eta' (u_1)\eta'(u_2)}{4J^2 \cos^2 \frac{\eta_R(u_1)-\eta_L(u_2)}{2}} \right)^\Delta \delta^{ij}.
\ee
Here, $\eta_L$ and $\eta_R$ are the restriction of AdS$_2$ global time $\eta$ to the left and right boundaries. At low energy limit, we expect the effective action for the boundary to take the following form:
\be
\begin{aligned} \label{newaction}
 &S = - \bigg[ \int \, d\tilde u \, Sch( \tan \frac{\eta_L}{2},\tilde u)+ \int \, d \tilde u \, Sch( \tan \frac{\eta_R}{2}, \tilde u)\bigg]-2\kappa \, \int \, d \tilde u \,  \bigg(\frac{\eta'_L(\tilde u) \, \eta'_R(\tilde u)}{ \cos^2 \frac{\eta_R-\eta_L}{2}}\bigg)^{\Delta} \\
&~~~~~~~~~~~~~~~~~~~\kappa = \frac{g}{2}(\frac{b}{(2J)^{2\Delta}})~ (\frac{\phi_r}{8\pi G})^{1-2\Delta} ,~~~~\tilde u = \frac{8\pi G u}{\phi_r},
\end{aligned} 
\ee
where we approximated the interaction term  by $S_{int} \approx \int \, du \, g(u) \, \blangle \,  \phi^i_L(u) \phi^i_R(u) \brangle$. We can rewrite the above action as:
\be
\begin{aligned}
&S = \int \, d \tilde u \, \bigg( P_{\eta_R} \eta'_R + P_{\eta_L} \eta'_L+ P_{\phi_R} \phi'_R+P_{\phi_L} \phi'_L - H \bigg), \\
&H = \frac{1}{2}\bigg(P^2_{\phi_L}+ e^{2\phi_L}+2P_{\eta_L} e^{\phi_L}\bigg)+\frac{1}{2}\bigg(P^2_{\phi_R}+ e^{2\phi_R}+2P_{\eta_R} e^{\phi_R}\bigg) +2\kappa \, \bigg(\frac{e^{ (\phi_R+\phi_L)}}{\cos^2 \frac{\eta_R-\eta_L}{2}} \bigg)^{\Delta}.
\end{aligned}
\ee
The equation of motion is given by:
\be \label{solution2}
\begin{aligned} 
&\phi'_R = P_{\phi_R},~~~~ \eta'_R = e^{\phi_R}, \\
&P'_{\phi_R} = - \bigg(e^{2\phi_R}+P_{\eta_R} e^{\phi_R}+2\kappa \Delta \bigg(\frac{e^{ (\phi_R+\phi_L)}}{\cos^2 \frac{\eta_R-\eta_L}{2}} \bigg)^{\Delta} \bigg), \\
& P'_{\eta_R} = -2\kappa \Delta \tan \frac{\eta_R - \eta_L}{2} \bigg(\frac{e^{ (\phi_R+\phi_L)}}{\cos^2 \frac{\eta_R-\eta_L}{2}} \bigg)^{\Delta}.
\end{aligned}
\ee
Note that \ref{newaction} still has SL(2,R) symmetry. The corresponding conserved charges are:
\be \label{noethers}
Q_3 = Q_3^R+Q_3^L, ~~~~ Q_2 = Q^R_2 - Q^L_2, ~~~~ Q_1 = Q^R_1 - Q^L_1.
\ee  
where $Q^{R(L)}_i$s are the charges \ref{noether}. Here, the Poisson bracket is with respect to  $(\phi_R,P_{\phi_R})$, $(\phi_L,P_{\phi_L})$, $(\eta_L,P_{\eta_L})$, and $(\eta_R,P_{\eta_R})$. Since we are interested in solving \ref{solution2} with initial condition set by \ref{TFD} for both sides, and both are symmetric with respect to left and right, $\eta_L(u) = \eta_R(u)$ is always guaranteed, so we can reduce the equation of motion to: 
\be \label{solution3}
\begin{aligned}
&\phi'_R = P_{\phi_R},~~~~ \eta'_R = e^{\phi_R}, ~~~~  P_\eta =0, \\
&\phi''_R = \begin{cases} - (e^{2\phi_R}+2\kappa   \Delta \, e^{2\Delta \phi_R}) ~~~~ u \geq 0  \\
- e^{2\phi_R} ~~~~~~~~~~~~~~~~~~~~~~~~\, u<0.
\end{cases}
\end{aligned}
\ee
One can solve the above equation for $u \ge 0$ perturbatively assuming that $\kappa \ll 1$ with the initial condition
\be
\phi(0) = \ln \frac{S}{2\pi},~~~~ \phi'(0) = 0.
\ee
In particular, we are interested in computing the Casimir function $Q_R = \phi_R'^2 + e^{2\phi_R} = Q_0 + \kappa Q_1 + \kappa^2 Q_2 + O(\kappa^3)$. Equation \ref{solution3} implies:
\be \label{Energyconserv}
{\phi'}_R^2 + e^{2\phi_R}+ 2\kappa e^{2\Delta\phi_R} = \Big(\frac{S}{2\pi} \Big)^2 + 2\kappa\Big(\frac{S}{2\pi} \Big)^{2\Delta}. 
\ee
Taking $\phi = \phi_0 + \kappa \phi_1 + \kappa^2 \phi_2 + O(\kappa^3)$ and plugging into \ref{solution3}, at leading order we have:
\be
\begin{aligned} &
\phi_0(\tilde u) = -\ln \frac{2\pi}{S} \cosh \Big( \frac{S}{2\pi} \tilde u \Big), ~~~~~~~Q_0 = \Big(\frac{S}{2\pi}\Big)^2, 
\end{aligned}
\ee
while in the next order one gets:
\be
\phi_1' -\frac{S}{\pi} \frac{1}{\sinh \frac{S \tilde u}{\pi}} ~\phi_1 = \Big(\frac{S}{2\pi} \Big)^{2\Delta-1} \Big( 1 - \frac{1}{\cosh^{2\Delta} \frac{S \tilde u}{2\pi}} \Big) \coth \frac{S \tilde u}{2\pi}.
\ee
One can rewrite the above as follows:
\be
 \Big(\coth \frac{S \tilde u }{2\pi}  \phi_1 \Big)' = \Big(\frac{S}{2\pi} \Big)^{2\Delta-1} \Big( 1 - \frac{1}{\cosh^{2\Delta} \frac{S \tilde u}{2\pi}} \Big) \coth^2 \frac{S \tilde u}{2\pi},
\ee
with the answer:
\be
\phi_1 = \Big(\frac{S}{2\pi} \Big)^{2\Delta-2} \bigg( - \frac{S\tilde u}{2\pi} \tanh \frac{S\tilde u}{2\pi} + 1 - \frac{1}{\cosh^{2\Delta} u} + (1-2\Delta) \tanh \frac{S \tilde u}{2\pi} \int_0^{\frac{S\tilde u}{2\pi}} \frac{1}{\cosh^{2\Delta} x} \Big).
\ee
The leading correction to the Casimir is given by:
\be
\begin{aligned} &
Q_1 =  2\bigg( \Big(\frac{S}{2\pi} \Big)^{2\Delta} - e^{2\Delta\phi_0}\Big) \bigg) =2 \Big(\frac{S}{2\pi} \Big)^{2\Delta} \bigg( 1 - \frac{1}{\cosh^{2\Delta} \frac{S~\tilde u}{2\pi}} \bigg), \\&
Q_2 = -4\Delta ~ e^{2\Delta \phi_0} ~ \phi_1.
\end{aligned}
\ee
Then the value of the entropy can be rederived from 
\be \label{casimir-entropy}
S(u) = 2\pi \sqrt{Q_R} = 2\pi \sqrt{Q_0} \bigg( 1 +\frac{\kappa}{2} \frac{Q_1}{Q_0} + \frac{\kappa^2}{2} \Big( \frac{Q_2}{Q_0} - \frac{1}{4} \Big( \frac{Q_1}{Q_0} \Big)^2 \Big) + O(\kappa^3) \bigg).
\ee
\begin{figure}[h]

\begin{subfigure}{0.3\textwidth}
\centering
\includegraphics[width=0.4\linewidth, height=8cm]{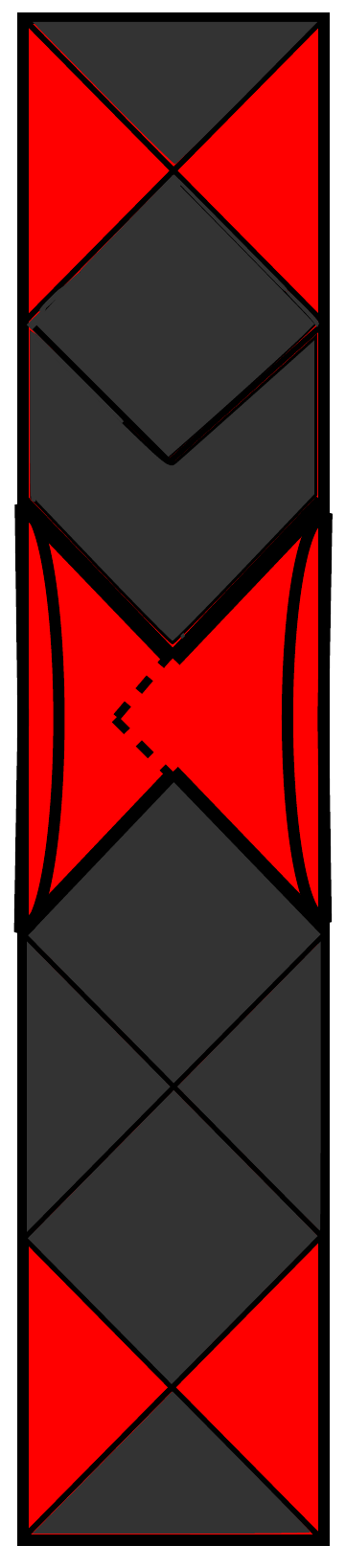} 
\caption{}
\label{fig:Disentangling}
\end{subfigure}
\begin{subfigure}{0.3\textwidth}
\centering
\includegraphics[width=0.4\linewidth, height=8cm]{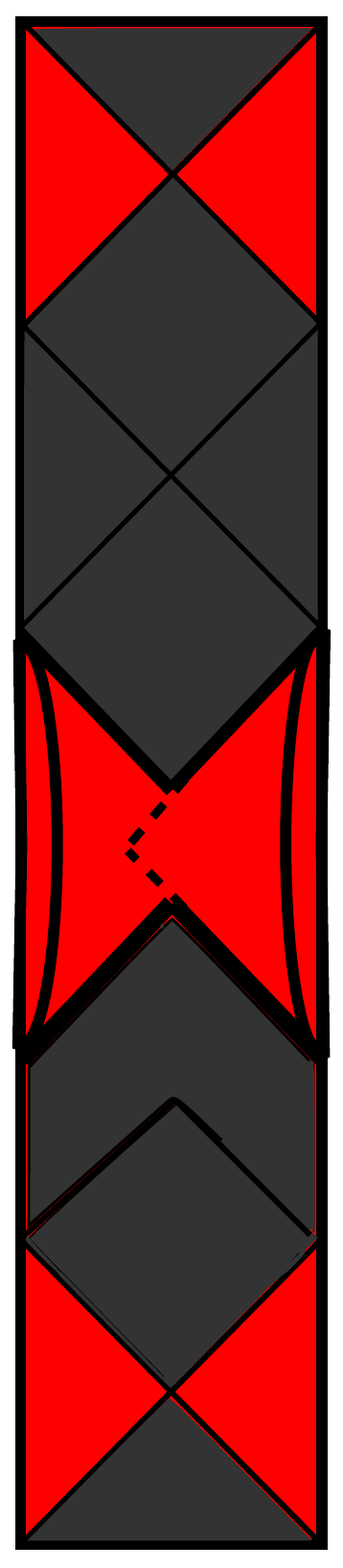}
\caption{}
\label{fig:timereverse}
\end{subfigure}
\begin{subfigure}{0.3\textwidth}
\centering
\includegraphics[width=0.4\linewidth, height=8cm]{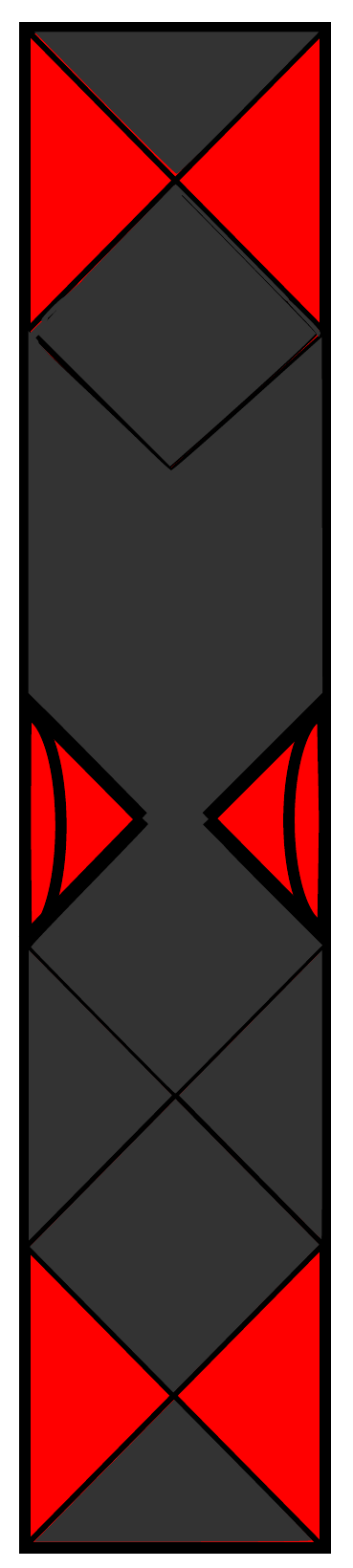}
\caption{}
\label{fig:entangling}
\end{subfigure}
\caption{The effect of the double trace operator on the black hole in the bulk. (a) represents the case where the interaction is on for $t>0$. For $g<0$ , the operator will decrease the entropy of the black hole, the black hole's size gets smaller, and so part of the region behind the horizon is now revealed to the outside observer. (b) displays the case where the interaction is on while we go backward in time which shrinks the white hole region. (c) We can repeat (a) but with $g>0$. This leads to expanding the black hole. The entanglement entropy will increase, the black hole will expand, so the red region will become smaller.  }
\label{fig:Dis}
\end{figure}
Considering the expansion $S(u) = S_0 + \kappa S_1(u) + \kappa^2 S_2(u)$, we have:
\be \label{entropy12}
\begin{aligned} &
S_1(u) = 2\pi \Big(\frac{S}{2\pi} \Big)^{2\Delta-1} \bigg( 1 - \frac{1}{\cosh^{2\Delta} \frac{S~\tilde u}{2\pi}} \bigg),~~~~~~~~~~~~\frac{S\tilde u}{2\pi} = \frac{2\pi u}{\beta}. \\&
S_2(u) = \frac{S}{2} \Big(\frac{S}{2\pi} \Big)^{4\Delta-4}\bigg( -\bigg( 1 - \frac{1}{\cosh^{2\Delta} \frac{S~\tilde u}{2\pi}} \bigg)^2 +\frac{4\Delta}{\cosh^{2\Delta} \frac{S~\tilde u}{2\pi}} \bigg( \frac{S \tilde u}{2\pi} \tanh \frac{S \tilde u}{2\pi}- 1 + \frac{1}{\cosh^{2\Delta} \frac{S\tilde u}{2\pi}}\\&~~~~~~~~~~~~~~~~~~~~~~~~~~~~~~~~~~~~~~~~~~~~~~~~~~~~~~~~~~~~~~~~~~~~~~~~~~~~~~-(1-2\Delta)\tanh \frac{S \tilde u}{2\pi} \int_0^{\frac{S\tilde u}{2\pi}} \frac{1}{\cosh^{2\Delta} x}\bigg)\bigg).
\end{aligned}
\ee
Note that $\kappa S_1(u)$ and $\kappa^2 S_2(u)$ are exactly equal to \ref{deltaent} and \ref{entropy2}. Notice that the relation of entanglement entropy to the coarsed-grained quantity \ref{casimir-entropy}, is more than the equality of final answers. In fact, in the second order, the term $\pi \kappa^2\sqrt{Q_0} \frac{Q_2}{Q_0}$ is equal to the second configuration in section \ref{secondorderentang}, and $-\frac{\pi \kappa^2}{4} \sqrt{Q_0} \Big( \frac{Q_1}{Q_0} \Big)^2$ equals the first configuration.
\subsection{The explicit solution for $\Delta = \frac{1}{2}$}
We can find an exact solution to \ref{solution3} for $\Delta = \frac{1}{2}$:
\be \label{trajectory}
\tan \frac{\eta}{2} = \frac{S}{\tilde S} ~ \tanh \frac{\pi u}{\tilde \beta},~~~~\tilde \beta = \beta ~\Big( 1+ \frac{4\pi \kappa}{S} \Big)^{\frac{-1}{2}},
\ee
and $\tilde S$ is the thermal entropy associated with the new inverse temperature $\tilde \beta$, $\Big( \frac{S}{\tilde S} = \frac{\tilde\beta}{\beta}\Big)$.
Plugging \ref{trajectory} into the formula for the Casimir yields:
\be
\tilde S(u) = 2\pi \sqrt{Q_R} = S \,\,\Bigg( \frac{1+(\frac{\tilde S}{S})^2 \tanh^2 \frac{\pi u}{\tilde \beta}}{1+(\frac{S}{\tilde S})^2 \tanh^2 \frac{\pi u}{\tilde \beta}} \Bigg)^{\frac{1}{2}}.
\ee
One can argue that the effect of the double traced operator for $g<0$ is to shrink the black hole, the Einstein-Rosen bridge, so part of the region that was behind the horizon is now revealed to the outside observer. Therefore,  for the black hole interior to be revealed to the outside observer, the observer has to decrease the entanglement entropy of TFD and reach the states  $\ket{\widetilde{\TFD}(t)}$ in the Hilbert space. As a consequence, the wormhole between the two sides will shrink, and part of the interior can be probed from outside; see Figure \ref{fig:Dis}. To quantify this, we define the length of the wormhole at time $t$ as the length of the geodesic that connects the two points at $r = \frac{1}{S}$. Shrinking a given wormhole corresponds to a configuration where the upper bifurcation point will be placed on the wormhole, which happens when we decrease the entanglement entropy. For example, for $\Delta = \frac{1}{2}$, the location of the upper bifurcation point as a function of $\kappa$ is:
\be
\Big(\eta_b , \sigma_b \Big) = \Big(2 \arctan \frac{1}{\sqrt{1-\frac{4\pi \kappa}{S}}}- \frac{\pi}{2},0\Big).
\ee
Now, to the leading order in $\frac{1}{S}$, the length of the wormhole satisfies:
\be
\cosh d_w = 1 + 2 \tan^2 \eta_b,
\ee
or equivalently:

\begin{figure}[t]
\centering
\includegraphics[scale=.3]{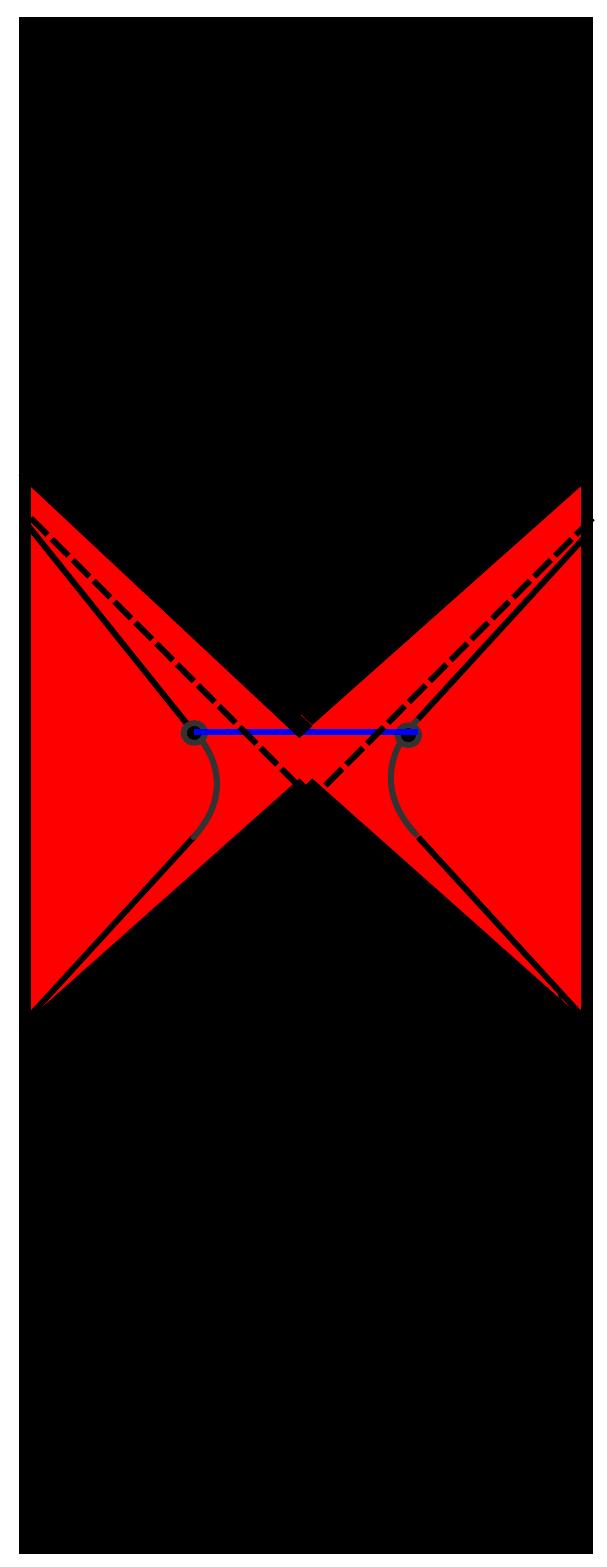}
\caption{The two-sided black hole embedded in the global AdS. The blue segment is the wormhole that connects the two sides whose endpoints are located on $r = \frac{1}{S}$ ( $r$ is the Schwarzschild coordinate). We can shrink such a wormhole by spending the entanglement that exists in the thermofield-double state.}
\label{wormholec}
\end{figure}
\be \label{shrinkage}
\frac{d_w}{2} = -\log \sqrt{1+\frac{4\pi \kappa}{S}} \approx \frac{2\pi |\kappa|}{S}+ O(\frac{\kappa^2}{S^2}),
\ee
where $d_w$ is the  length of the wormhole. \ref{shrinkage} implies that in order to shrink a wormhole of  length $d_w$, one needs to decrease the entanglement entropy to $S-2\pi |\kappa |$. This may be a manifestation of ER=EPR \cite{MS13}. To send a message from one side to the other side, we also need to take into account the back reaction of the message on the black hole, which leads to the expansion of the wormhole. Such backreactions are due to the shock wave effects \cite{DtH85, Shenker14}. Therefore, to send a message between  two parties with a black hole in between, in general, there are two barriers: the wormhole that already exists and the back reaction of the message that leads to its expansion. However, one can use the entanglement as a resource to shrink the wormhole and send the message to the other side.  \\
We can also interpret \ref{Energyconserv} as the conservation of the energy for a particle whose trajectory is the boundary of AdS:
\be 
{\phi'}_R^2 + e^{2\phi_R}+ 2\kappa e^{2\Delta\phi_R} = E = \Big(\frac{S}{2\pi}\Big)^2 + 2\kappa\Big(\frac{S}{2\pi} \Big)^{2\Delta}. 
\ee
This implies that at late times we have:
\be
S(u \rightarrow \infty) = 2\pi \sqrt{E} =  S ~ \sqrt{1 + 2\kappa \Big(\frac{S}{2\pi} \Big)^{2\Delta-2}  }.
\ee
This implies a critical value for the coupling $\kappa$ where the energy of the particle (the entanglement entropy) vanishes:
\be
\kappa_* = \frac{-1}{2} \Big(\frac{S}{2\pi}\Big)^{2-2\Delta}
\ee
Naively, $\kappa_*$ is the value for which the state's thermal entropy vanishes. However, close to this value, our classical computation becomes invalid ($\frac{\phi_r}{\beta J} \sim 1$), and one should consider the quantum Schwarzian \cite{Kisuh18B,BAK16b,MH17a,MH17b}. \\
\section{Conclusion}
In this article we studied the evolution of the thermofield-double at temperature $\beta$, $\ket{\TFD ; \beta}$, by double-traced operators connecting its both sides. The resulting state was denoted by $\ket{\widetilde \TFD (t)}$. Based on our computation, there is a strong evidence that for holographic theories at $t \gtrsim  \beta$ we have:
\be \label{tfdtilde}
\ket{\widetilde\TFD(t)} \approx \ket{\TFD; \tilde \beta}.
\ee
where $\tilde\beta$ is given by \ref{cgtemp}.
We showed this in our one-dimensional theory by observing that, to leading order in the coupling, the entanglement entropy and the coarse-grained entropy will remain equal. The key point is that at low temperature, the unperturbed theory will be described by the Schwarzian modes where the connected part of the time-ordered four-point function has the following simple form \cite{MS16}:
\be
\calF_{TO} \Big(\theta_1,\theta_2,\theta_3,\theta_4\Big) = \frac{\delta_{\beta} G}{G} (\theta_1, \theta_2) \frac{\delta_{\beta} G}{G} (\theta_3,\theta_4) ~ \langle \delta \beta ^2 \rangle.
\ee
Furthermore, we computed the Casimir function associated with the Noether charges of the effective action and observed that its value exactly matches the result of our microscopic computation. Since the Casimir is a coarse-grained quantity, one would expect that the equality between the entanglement and coarse-grained entropies remains valid at higher orders in the coupling. \\\\
Our results also shed light on the geometry of a two-sided black hole in the bulk. The left and right wedges are the representation of the thermofield-double. While the two wedges are causally disconnected, they are ``correlated'' due to the entanglement. The other two wedges, the black hole and the white hole, are not accessible from the outside. To reach the inside, one should have access to less entangled states. One way is to use the interaction \ref{Hamilint} to decrease the entropy. Consequently, according to an outside observer, a particle, thrown into the bulk, never reaches the event horizon in finite time; this corresponds to a local perturbation of the thermofield-double, and so not only does not it change the entanglement entropy, it, in fact, increases the $\TFD$'s coarse-grained entropy and so the black hole's size, as a result. \\\\
It is important to see whether \ref{tfdtilde} holds in $d \ge 4$.  In general, when a system reaches equilibrium, one expects its coarse-grained entropy to be bigger than its entanglement entropy. However, for holographic theories, one may expect the interaction \ref{Hamilint} to act coherently, and so \ref{entanglement change} is the coarse-grained entropy change. This would  mean that a holographic theory has a ``gravitational sector'' whose modes are, for example, responsible for the saturation of the chaos bound\cite{MSS}.
\\\\ \large{\bf Acknowledgements}
\normalsize
\\
I am grateful to J. Maldacena, D. Stanford, H. Verlinde for helpful discussions. I am grateful to J. Suh for making a comment on the draft of the paper, to J. Preskill for insightful discussions, and especially to A. Kitaev for helpful discussions and comments at different stages of this project, without which I would not be able to finish the paper. I also acknowledge support from the Simons Foundation \,(award number 376205).

\bibliography{Traversable}
\bibliographystyle{JHEP}
\end{document}